\documentclass[twocolumn]{aastex61hack}

\usepackage{natbib}
\newcommand{\HI}{\ensuremath{\mbox{\ion{H}{1}}}}
\newcommand{\HII}{\ensuremath{\mbox{\ion{H}{2}}}}

\newcommand{\NII}{\ensuremath{[\mbox{\ion{N}{2}}]}}
\newcommand{\SII}{\ensuremath{[\mbox{\ion{S}{2}}]}}

\newcommand{\Ha}{\ensuremath{\rm H\alpha}}
\newcommand{\halpha}{\ensuremath{\rm H\alpha}}

\newcommand{\htwo}{\ensuremath{{\rm H}_2}}

\newcommand{\logNHI}{\ensuremath{\log N(\mbox{\ion{H}{1}})}\relax}

\newcommand{\z}{$z$}

\newcommand{\msun}{M$_\odot$}

\newcommand{\msunyr}{M$_\odot$~yr$^{-1}$}

\newcommand{\percc}{cm$^{-3}$}

\newcommand{\kms}{km~s$^{-1}$}

\newcommand{\wfpc}{WFPC2}
\newcommand{\hst}{{\em HST}}

\newcommand{\ngc}{NGC~}

\newcommand{\epsO}{\ensuremath{\epsilon ({\rm O})}}
\newcommand{\DeltaO}{\ensuremath{\Delta \epsilon ({\rm O})}}

\newcommand{\lhalpha}{\ensuremath{L_{{\rm H}\alpha}}\relax}

\defcitealias{howk2018a}{Paper~I}

\begin{document}

\title{Extraplanar \HII\ Regions in Spiral Galaxies. II. In Situ Star Formation in the Interstellar Thick Disk of NGC 4013}


\author[0000-0002-2591-3792]{J. Christopher Howk}
\affiliation{Department of Physics, University of Notre Dame,
  Notre Dame, IN 46556, USA}
\affiliation{Instituto de Astrof\'{i}sica,
  Pontificia Universidad Cat\'{o}lica de Chile, Santiago, Chile}
\author{Katherine M. Rueff}
\affiliation{Department of Physics, University of Notre Dame,
  Notre Dame, IN 46556, USA}
\author{Nicolas Lehner}
\affiliation{Department of Physics, University of Notre Dame,
  Notre Dame, IN 46556, USA}
\author{Christopher B. Wotta}
\affiliation{Department of Physics, University of Notre Dame,
  Notre Dame, IN 46556, USA}
\author{Kevin Croxall}
\affiliation{Department of Astronomy, The Ohio State University,
  Columbus, OH 43210, USA}
\affiliation{Illumination Works LLC, 5550 Blazar Parkway \#150, Dublin,
   OH 43017, USA}
\author{Blair D. Savage}
\affiliation{Department of Astronomy, University of Wisconsin,
  Madison, Madison, WI 53706, USA}

\begin{abstract}
  We present observations of an \halpha\ emitting knot in the thick disk of
  NGC~4013, demonstrating it is an \HII\ region surrounding a cluster of young
  hot stars $z = 860$ pc above the plane of this edge-on spiral galaxy. With
  LBT/MODS spectroscopy we show this \HII\ region has an \halpha\ luminosity
  $\ge 4$ - 7 times that of the Orion nebula, with an implied ionizing photon
  production rate $\log Q_0 \ga 49.4$ (photons s$^{-1}$). \hst /WFPC2 imaging
  reveals an associated blue continuum source with $M_{V} = -8.21\pm0.24$.
  Together these properties demonstrate the \HII\ region is powered by a young
  cluster of stars formed {\em in situ} in the thick disk with an ionizing
  photon flux equivalent to $\sim$6 O7 V stars. If we assume $\approx6$ other
  extraplanar \halpha -emitting knots are \HII\ regions, the total thick disk
  star formation rate of \ngc 4013 is $\sim 5 \times 10^{-4}$ M$_\odot$
  yr$^{-1}$.  The star formation likely occurs in the dense clouds of the
  interstellar thick disk seen in optical images of dust extinction and CO
  emission.
\end{abstract}



\section{Introduction}

Stars form in cold, dense molecular clouds, from cores likely shaped by
supersonic turbulence that collapse under the influence of gravity
\citep{mckee2007}.  The primary physical conditions needed for star formation
are largely confined to the high-density regions in galactic disks. However,
star formation has also been discovered in unusual locations throughout the
universe, including in the far outer disks of galaxies
\citep[e.g.,][]{ferguson1998, werk2008, thilker2007, barnes2011}, in
tidally-stripped gas \citep[e.g.,][]{de-mello2012, stein2017,
urrutia-viscarra2017}, in ram-pressure stripped material \citep{cortese2004,
kenney2004}, and even in a potentially-primordial ring of gas surrounding a
group of galaxies \citep{thilker2009}. These cases offer the opportunity to
study the drivers of star formation in differing conditions.

Another extreme location for star formation is in the thick disks of spiral
galaxies, where the pressures are much lower than in the midplane. Interstellar
thick disks (within $\approx \pm 2$ kpc, where the gas mosty corotates with the
thin disk) are multiphase, including substantial quantities of dense or even
molecular gas \citep{garcia-burillo1999, howk1999, howk2000}.  {\em Candidate}
thick disk \HII\ regions have been identified in a handful of galaxies as knots
of extraplanar \Ha\ emission \citep{walterbos1991, rand1996, howk2000,
rossa2000, cortese2004, rueff2013}.  \citet{tullmann2003} presented the first
spectroscopic confirmation of two \HII\ regions in the thick disk/halo of \ngc
55 ($z\approx$1.1 kpc, 2.2 kpc), which are consistent with being powered by
single late-O/early-B stars.

This paper is the second in a series to study the implications of extraplanar
\HII\ regions in nearby edge-on galaxies. In the first of the series
\citep[][hereafter \citetalias{howk2018a}]{howk2018a} we considered the emission
line spectrum from the brightest extraplanar \HII\ region candidate identified
by \citet{rueff2013} at $z = 860$ pc above the plane of NGC~4013 (at
$[\alpha,\delta]_{\rm J2000} = $[11:58:33.2, +43:57:11.85]).\footnote{We adopt
$D=17.1\pm1.7$ Mpc to NGC~4013 (distance modulus of $31.17\pm0.10$ mag), the
mean distance to galaxy group 102 within which NGC~4013 resides
\citep{tully2008}.} We demonstrated that the knot of \Ha\ emission was indeed an
\HII\ region associated with \ngc 4013. This nebula, \ngc 4013 EHR1 (EHR =
extraplanar \HII\ region following \citealt{tullmann2003}) is projected
$R\approx 2.5$ kpc radially from the nucleus and $z = 860$ pc above the
plane. We showed in \citetalias{howk2018a} that its metallicity is $\approx1/2$
that of \HII\ regions in the disk of \ngc 4013 and discussed the implications of
this low metallicity for flows through the thick interstellar disks of galaxies.

\begin{figure}
  \epsscale{1.15}
\plotone{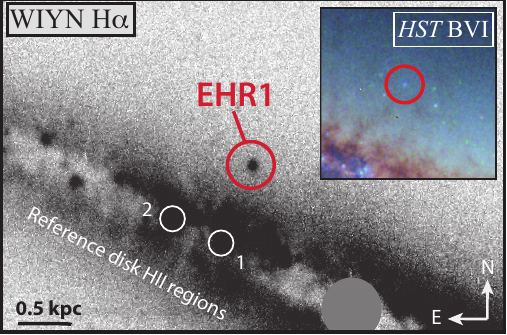}
\caption{WIYN \Ha\ image (greyscale) and \hst\ $BVI$ image (three-color inset)
of the brightest extraplanar \HII\ region candidate in \ngc 4013 \citep[images
described in][]{rueff2013}. The extraplanar \HII\ region \ngc 4013 EHR1 lies at
$z = 860$ pc above the midplane. MODS longslit spectra reveal it is an \HII\
region, while the \hst\ photometry implies a cluster of young, O-type stars as
the ionizing source for the nebula.  Also marked are the two disk \HII\ regions
used to define the disk reference abundance \citepalias[see][]{howk2018a}. A
bright foreground star is covered by a gray disk near the bottom of the image.
\label{fig:image}}
\end{figure}

In this paper we focus on the nature of the stars that are powering the nebula
\ngc 4013 EHR1. Figure \ref{fig:image} shows WIYN \Ha\ and {\em Hubble Space
Telescope} (\hst)/WFPC2 broad-band imagery \citep{rueff2013} of the nebula and
its vicinity. The \hst\ images show a blue continuum source coincident with the
\Ha\ emission. In what follows we demonstrate that the continuum emission is
inconsistent with emission from a single star, implying it is a cluster, and
that the evolutionary timescales imply the cluster was formed {\em in situ} in
the thick disk. Our paper is arranged as follows. In \S  \ref{sec:spectroscopy}
we briefly discuss the spectroscopy of this object and the implications for the
underlying source that can be derived from the parameters of the \HII\ region.
In \S \ref{sec:photometry} we discuss the \hst\ imaging of \cite{rueff2013},
deriving new photometric properties of the continuum emission associated with
EHR1 and the origins of this emission. Lastly we discuss the implications of
{\em in situ} star formation in the thick disks of galaxies in \S
\ref{sec:discussion} and summarize our results in \S \ref{sec:summary}.

\section{Emission Line Spectroscopy of the Extraplanar Nebula
  NGC~4013 EHR1}
\label{sec:spectroscopy}

\subsection{LBT/MODS Observations}

We observed the \Ha\ knot EHR1 using one side of the 8.4-m Large Binocular
Telescope equipped with the first of the Multi-Object Double Spectrographs
(MODS).\footnote{MODS is fully described in \citet{pogge2010}.} We obtained
$3\times1200$ sec longslit spectroscopy of NGC~4013 EHR1 using a $1\farcs0$ wide
longslit with coverage $3,200 \la \lambda \la 10,000$ \AA. We use the MODS
pipeline\footnote{The MODS reduction pipeline was developed by K. Croxall with
funding from NSF Grant AST-1108693. Details can be found at {\tt
http://www.astronomy.ohio-state.edu/MODS/Software/modsIDL/}.} to reduce the
spectra, with details in \citetalias{howk2018a}.

\begin{figure*}
\epsscale{1.2}
\plotone{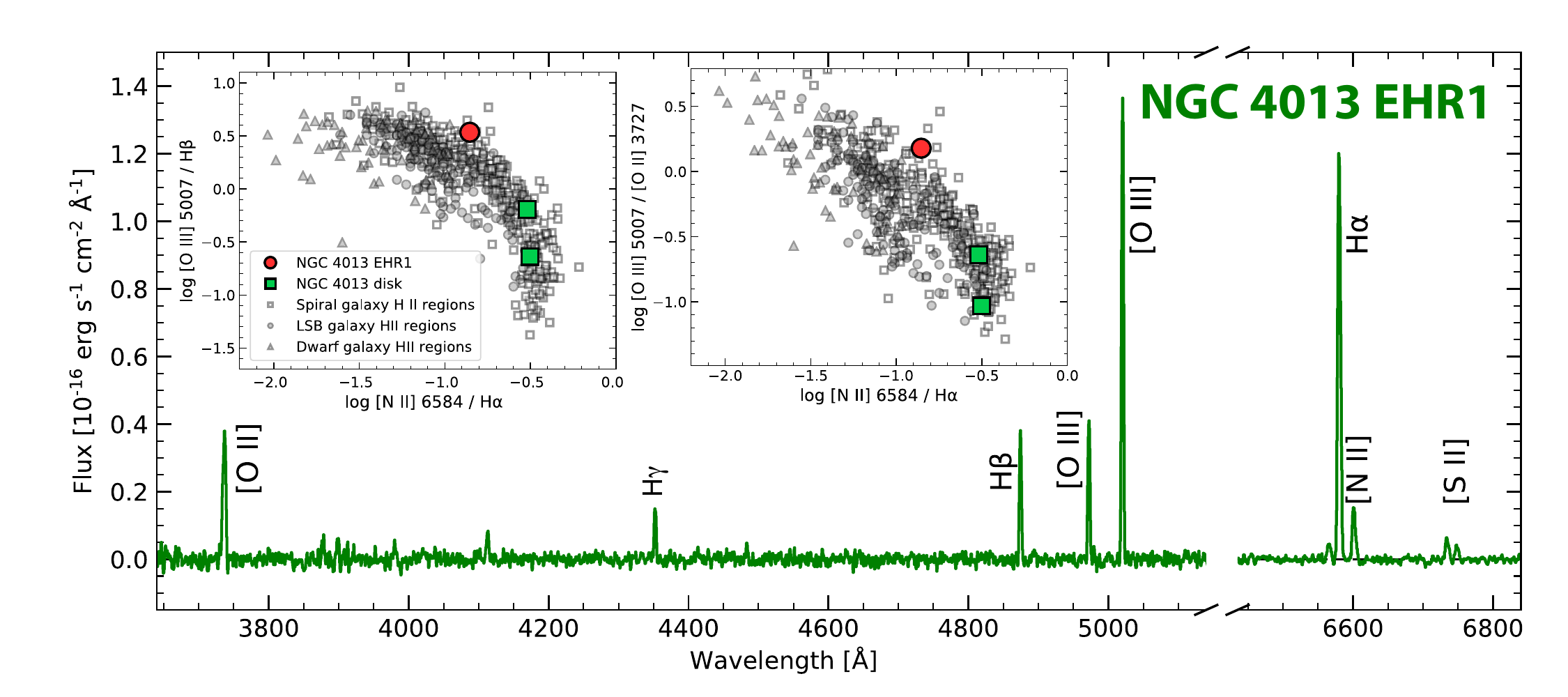}
\caption{LBT/MODS spectrum of \ngc 4013 EHR1. Several hydrogen Balmer lines and
collisionally-excited metal lines are shown, notably those important for
assessing the gas-phase abundance. {\em Inset} -- Characteristic emission line
diagnostic ratios for \ngc 4013 EHR1 compared with \HII\ regions in the disks of
spiral, low surface brightness, and dwarf galaxies in the local universe
\citep{berg2015, croxall2015, croxall2016, bresolin2015, van-zee2006}. EHR1 has
emission line ratios consistent with those seen in other high-excitation \HII\
regions in the local universe.  \label{fig:spectrum}}
\end{figure*}

Our final LBT/MODS spectrum is shown in Figure \ref{fig:spectrum}, and a summary
of derived properties important for the nature of the underlying stars is given
in Table \ref{tab:results}. The emission line intensities are available in
\citetalias{howk2018a}. In deriving the properties of the nebula EHR1 we have
subtracted a background from the local stellar and diffuse ionized gas (DIG)
emission. The DIG is particularly important, as it adds an emission line
background with line ratios quite different than those of the nebula
\citep{haffner2009}. Our background spectrum is based on the geometric
mean of regions immediately adjacent to either side of our extraction box for
EHR1 (as we assume the emission falls off exponentially). We use a $1\farcs5$
extraction window, so that the DIG background is sampled $0\farcs75$ to
$2\farcs25$ from the center of EHR1. Thus the DIG is sampled 36 to 100 pc from
the center of our target.

While the subtraction of the background DIG emission can have a sizeable impact
on the line ratios used for estimating the metallicity, it has only a small
impact on the diagnostics that impact our derived properties of the underlying
stars (the subject of this paper). Our main diagnostics of the underlying stars
are the Balmer lines, in which the DIG is significantly fainter than EHR1. Our
DIG background spectrum has an $\halpha$ intensity that is only $\approx13\%$ of
the emission from EHR1 (whereas the intensities of the forbidden \NII\ and \SII\
are nearly equal from the DIG and EHR1). The errors in the $\halpha$ luminosity
are more than twice this value (dominated by the extinction correction). Thus
errors in the background spectrum would have to be much larger than expected to
have significant impact on our total error budget.

The other main property derived from our spectroscopy that bears on the origin
of the gas and stars is the velocity of the nebula EHR1 (Table
\ref{tab:results}). The appropriateness of that quantity depends on the nebula
being placed accurately in the center of the slit. We positioned the slit using
a blind offset from a nearby bright star (with offsets derived from our WIYN
imaging). The pointing of the instrument in these conditions should be accurate
to a tenth of an arcsecond, implying a potential systematic error in the
velocity of the \HII\ region of up to $\approx25$ \kms. Velocities derived for
the DIG emission, which should largely fill the slit, should be limited only by
the wavelength calibration of the spectrograph (which we confirm with sky
absorption lines), though the intrinsic resolution is smaller given that it
fills the slit. In the end we adopt an error of $\approx \pm 25$ \kms\ for both.

\subsection{Characteristics of the Extraplanar \HII\ Region}

The \ngc 4013 EHR1 spectrum in Figure \ref{fig:spectrum} is typical of a
high-excitation \HII\ region (see insets) and inconsistent with the thick disk
DIG. The DIG near EHR1 has \NII /\Ha$\, \approx1$, whereas the nebula itself
shows a ratio $\approx0.1$.

We find a velocity $v_{\rm EHR1} = +720 \pm 25$ \kms, compared with a systemic
velocity for \ngc 4013 of $v_{\rm sys} = +835$ \kms\ \citep{bottema1995} (all
heliocentric). The DIG (in our spectra), \HI\ 21-cm emission
\citep{zschaechner2015}, and CO emission \citep{garcia-burillo1999} in this
direction all show gas at velocities consistent with EHR1. The nebula is clearly
associated with \ngc 4013 and at velocities consistent with the local thick disk
gas. The spectrum and velocity together demonstrate that \ngc 4013 EHR1 is an
\HII\ region in the thick disk of \ngc 4013.

Ideally one may be able to understand where EHR1 arises along the line of sight
with enough information about the velocity structure of the local extraplanar
material. The mean velocity of the DIG in our background spectra is $v_{\rm DIG}
= +680 \pm 25$ \kms, perhaps somewhat lower than EHR1. The local \HI\ emission
averaged over $\sim1$ kpc is peaked at $v \approx +785$ \kms, with emission
extending as low as $v \approx +650$ \kms\ \citep{zschaechner2015}. The
brightest \HI\ emission arises at lower \z\ than EHR1, though EHR1 lies well
within the thick disk emission seen in the cubes of
\citeauthor{zschaechner2015}.  In practice, assessing EHR1's relationship to
these ISM phases is complicated by the low velocity resolution of our DIG
measurements and the low spatial resolution of the \HI\ measurements.


In \citetalias{howk2018a} we found find a characteristic abundance $\epsO \equiv
12+\log ({\rm O/H}) = 8.35\pm0.15$ using the N2[$\equiv I(\NII \,
6584)/I(\halpha)$] scale of \cite{pettini2004}.  The thick disk \HII\ region has
an abundance lower than the disk by $\DeltaO = -0.32 \pm 0.09$, i.e., by a
factor of $\approx2$ \citepalias{howk2018a}.  In this case, the comparison is
made against the two \HII\ regions in the disk identified as references in
Figure \ref{fig:image}.  This discrepancy in metallicities of EHR1 compared with
the disk implies a flow of metal-poor gas through the thick disk in this galaxy.
It is unclear if this flow of matter is related to the formation of the cluster
itself (we will need metallicity measurements of more extraplanar \HII\ regions
to demonstrate this).

The physical properties of the nebula are summarized in Table \ref{tab:results}.
The extinction-corrected \Ha\ luminosity (see below) is $\lhalpha \approx
(4.0\pm1.2) \times10^{37}$ ergs s$^{-1}$ (errors dominated by the extinction
correction).  In deriving this value, we apply corrections for slit losses,
assuming a Gaussian image of 1\arcsec\ viewed through a perfect 1\arcsec\ slit,
requiring a factor of 1.3 correction to our total fitted intensity. This value
is $\ge (4 - 7) \times \lhalpha({\rm Orion})$\footnote{The Orion nebula
luminosity assumes a distance from \citet{sandstrom2007} and integrated flux
from \citet{rumstay1984}.} and requires an ionizing photon luminosity $\log
Q_{0} \ge 49.4$ \citep{osterbrock2006}. This is equivalent to $\sim$6
O7~V-equivalent stars \citep{martins2005}. Although the \Ha\ luminosity could
nearly be met by a single O4~V star, we rule this out in \S
\ref{sec:photometry}.

\begin{deluxetable}{lccr}
\tabletypesize{\scriptsize}
\tablenum{1}
\tablewidth{0pc}
\tablecolumns{2}
\tablecaption{Properties of NGC~4013 EHR1 \label{tab:results}}
\tablehead{Quantity & Value}
\startdata
%
\cutinhead{Stellar Properties}
{$L_{H_\alpha}$(ergs s$^{-1}$)} &
          $\ge (4.0\pm1.2) \times 10^{37}$\tablenotemark{a}\\
{log $Q_{0}$} &  $>49.3$ \\
{$M_{V}$} & $-8.21 \pm 0.24$\tablenotemark{a}\\
{$(B-V)_{0}$}  & $-0.36 \pm 0.10$\\
{$(V-I)_{0}$}  & $+0.06 \pm 0.16$\\
%
\cutinhead{\HII\ Region Gas Properties}
{$v_{\rm helio}$} (\kms)\tablenotemark{b} & $+720\pm25$ \\
{$T_{e}$ (K)}         & $\le 13200$\\
{$n_{e}$ (cm$^{-3}$)} &  $5 - 100$ \\
{12 + log (O/H)}     & $8.35 \pm 0.15$\\
{Metallicity ($Z_{\odot}$)\tablenotemark{c}} & $0.43 \pm 0.18 $\\
\enddata
\tablenotetext{a}{Assuming a distance of $17.1\pm1.7$ Mpc \citep{tully2008} and
an extinction given in the text. Errors include a contribution from the distance
uncertainties where appropriate.}
\tablenotetext{b}{Based on a joint fit to the Balmer line velocities. The
forbidden metal lines yield a similar velocity.}
\tablenotetext{c}{Assuming $12 + \log({\rm O/H})\odot = 8.72 \pm 0.03$
\citep{steffen2015}. The more secure result is that \ngc 4013 EHR1 has a
gas-phase oxygen abundance lower by $-0.32\pm0.09$ dex than the disk of \ngc
4013.}
\end{deluxetable}


\section{The Origins of the Stellar Cluster Associated with NGC~4013 EHR1}
\label{sec:photometry}

\subsection{\hst / WFPC2 Imaging of the Continuum Source}

The central cluster of NGC~4013 EHR1 is detected in the three-filter
(F435W/F555W/F814W) \hst /WFPC2 images of \cite{rueff2013}, as shown in Figure
\ref{fig:image}.  We derive Johnson-Cousins magnitudes from aperture photometry
$\{ B,V,I_C \}_{Vega} = \{ 23.51\pm0.04,23.65\pm0.06,23.24\pm0.11
\}$\footnote{We derive filter transformations assuming a mid-O star spectral
type calculated with the STSDAS routine {\tt calcphot}. The filter
transformations do not vary strongly with OB-star subtype ($<$0.02 mag).}  Our
Balmer line analysis implies $E(B-V) = 0.22 \pm 0.07$ -- equivalent to $A_{V} =
0.69 \pm 0.21$ and $E(V-I) = 0.35 \pm 0.11$ assuming a \citet{cardelli1989}
extinction curve and $R_V=3.1$. The foreground extinction due to the Milky Way
in this direction contributes $A_{V} \approx 0.05$ mag of the total
\citep{schlafly2011}; thus most of the extinction arises within \ngc 4013
itself.  Together our photometry and extinction estimates yield the intrisic
colors and absolute magnitude of the central source summarized in
Table~\ref{tab:results}.

\begin{figure}
  \epsscale{1.3}
\plotone{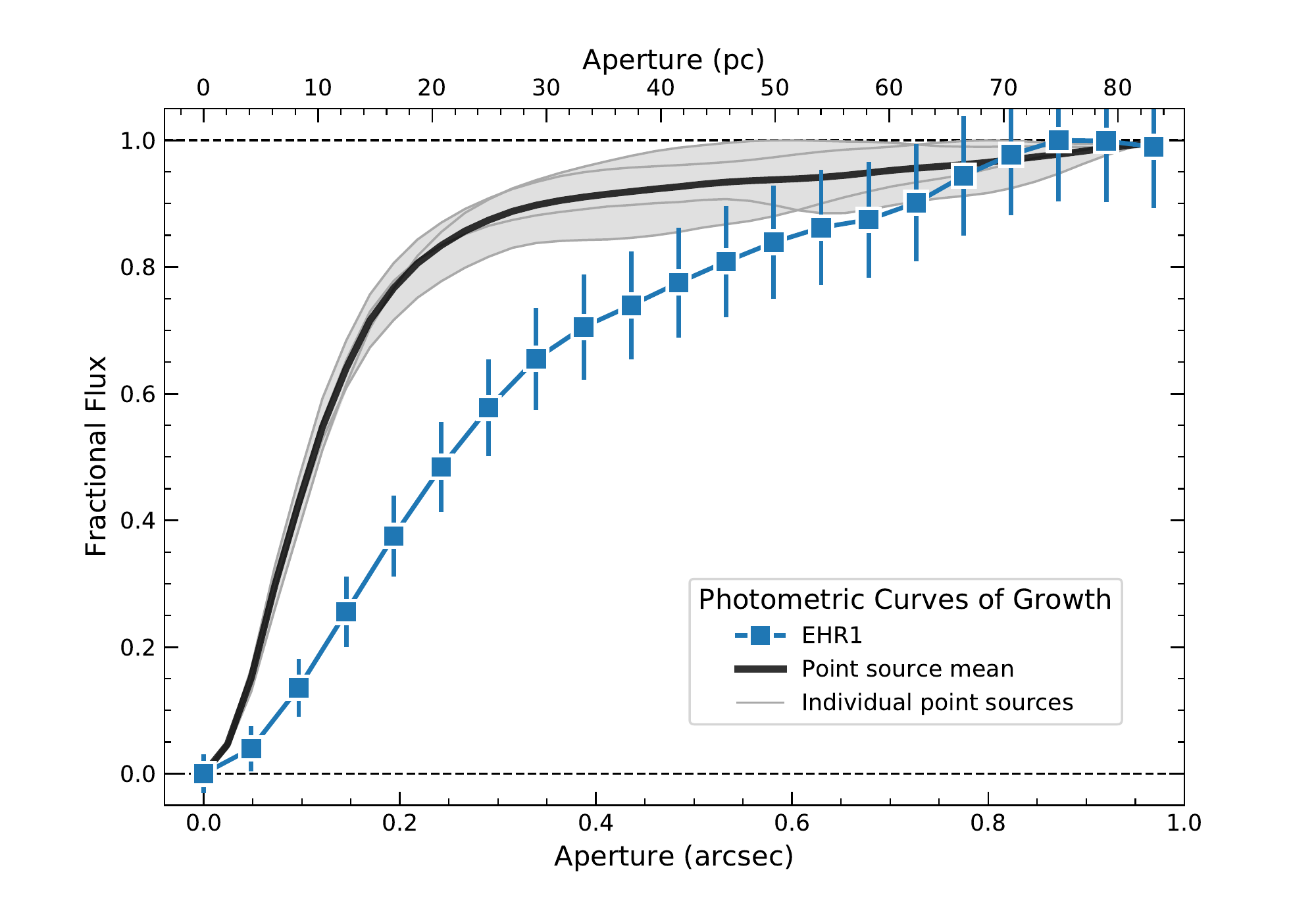}
\caption{The photometric curve of growth of EHR1 (blue) compared with the mean
behavior of three reference point sources (black) showing the fraction of the
integrated flux contained within apertures of differing sizes. The aperture
radii are given in arcseconds along the bottom and parsecs along the top of the
plot.  The shaded regions show the extrema from the three reference sources at
each aperture size, while the curves for the individual reference objects are
shown as thin grey lines. The light from EHR1 is clearly more extended than for
the reference sources. The uncorrected half-light radius of EHR1 is
$\sim0\farcs25$ or $\sim22$ pc.  Assuming the mean profile of the reference
sources is appropriate for deconvolving the EHR1 data, the intrinsic half-light
radius is $\sim0\farcs22$ or $\sim19$ pc. \label{fig:cog}}
\end{figure}

The \wfpc\ images demonstrate the continuum source underlying the extraplanar
\HII\ region is partly-resolved and thus not a single star. Figure \ref{fig:cog}
shows the photometric curve of growth (the amount of the light enclosed within
varying sized apertures) for the F555W (V-band) images of \ngc 4013 EHR1 (blue)
compared with the mean behavior of three likely point sources (thick black
line). For each source, we normalize the curves to their total integrated flux
within $1\farcs0$ (with sky subtraction performed locally based on an annulus
with inner and outer radii $1\farcs0$ and $1\farcs5$). The grey shaded area
shows the extreme values from our reference objects at each aperture size, with
the curves for the individual objects shown as thin grey lines. Our reference
sources were chosen to be {\em likely} point sources: identifying bona fide
point sources in the small WFPC2 field of view -- which is dominated by \ngc
4013 --  is difficult. It is made even more difficult by the presence of
clusters associated with \ngc 4013 itself. We vetted potential reference point
sources initially through visual inspection to rule out partially resolved
galaxies. We fit our final candidates with a two-dimensional Gaussian (though a
Moffat function fits as well), and we found the residuals for these three are
largely symmetrical (the fit gives no evidence for any ellipticity).  All four
reference sources are in very good agreement to $\sim75\%$ of the encircled
light, thus the core of the point spread function traced by these objects is
well characterized through the mean curve in Figure \ref{fig:cog}.  It is clear
from Figure \ref{fig:cog} that EHR1 is more extended than these sources.

The continuum source has a (deconvolved) FWHM$\, \approx12$ pc ($0\farcs144$)
and half-light radius $\approx19$ pc ($0\farcs22$) in the F555W images. The FWHM
values are based on two-dimensional Gaussian fits to the light profiles. We use
the mean FWHM of fits to the reference point sources to deconvolve the EHR1
fits. This size is on the large end -- but consistent with -- those of blue
clusters found in nearby galaxies \citep[e.g.,][]{larsen1999}.

\subsection{A Thick Disk Star Cluster Formed {\em in situ}}

Our \hst /\wfpc\ images demonstrate NGC~4013 EHR1 hosts a cluster of stars at
$z=860$ pc above the plane. In addition to being resolved, the source is too
bright to be a single star. We find $M_{V} = -8.21\pm0.24$, which is 2.7 and 3.6
magnitudes brighter than single O4~V or O7~V stars, respectively
\citep{martins2005}. The optical output is thus equivalent to $\sim10$ O4~V or
30 O7~V stars. The continuum emission is $\sim2$ magnitudes ($\sim 5 \times$)
brighter than even the brightest individual supergiants \citep{martins2005}.
Thus, the photometric properties of the source indicate the underlying continuum
source must represent many O stars. Following the Appendix of \citet{werk2008},
this number of O7 V-equivalent stars implies the presence of 5-200 O-type stars
(of all subtypes; this calculation depends on the highest-mass star present,
which is unknown). The ionizing photon production powering the \halpha\ emission
is dominated by the earliest spectral types, while the optical output has a
larger contribution from later spectral types. It could be possible that some of
the ionizing radiation is leaking from the \HII\ region into the neighboring
DIG, given the larger requirements of the optical emission compared with the
\halpha\ luminosity. However, this simple comparison does not completely account
for the full optical contributions from all the stars likely to be present (an
accounting that is made difficult by stochastic sampling of the initial mass
function in this relatively low-mass cluster).

The intrinsic colors for the stellar cluster associated with NGC~4013 EHR1 imply
the optical emission is dominated by O stars, as the colors are consistent with
the degenerate colors produced by these massive stars \citep{martins2006}.  If
the light from this object were dominated by an individual star, the colors
limit the age to $\tau \lesssim 3-10$ Myr \citep[][]{schaller1992}.  For the
integrated light from a cluster, the $(B-V)_0$ and $(V-I)_0$ colors are
indicative of ages $\tau \lesssim 6$ Myr \citep{pandey2010}. If this cluster was
ejected after forming in the disk of the galaxy, reaching its observed height in
this time would require a vertical velocity of $v_{z} \ga 100$ -- 300 \kms. This
is an extremely high peculiar velocity for a {\em cluster} of stars.  Even if
one could imagine a mechanism for providing this vertical kick to an entire
cluster of stars, if the vertical velocity is added to an assumed azimuthal
velocity due to rotation, the total cluster speed would approach or exceed the
escape speed of the galaxy.


It could be postulated that this thick disk cluster and nebula in \ngc 4013 are
associated with a disrupted dwarf galaxy. However, this is extremely unlikely,
as the accreted cluster would have to be deposited into the thick disk at an
extremely small -- and statistically-unlikely -- distance from the center of
\ngc 4013 \citep[the ``bulls-eye'' problem of ][]{peek2009}. Though there is
evidence for a major merger $3-5$ Gyr ago \citep{wang2015} based on \ngc 4013's
prodigious warp \citep{bottema1996, zschaechner2015} and recently-discovered
stellar stream \citep{martinez-delgado2009}, the very young age of the stars in
this cluster are inconsistent with the timescales of that event.

Thus, we conclude NGC 4013 EHR1 is powered by a cluster of young, hot
stars that formed \textit{in situ} within the thick disk of \ngc 4013.

\section{Thick Disk Star Formation}
\label{sec:discussion}

\ngc 4013 EHR1 is a thick disk \HII\ region located at $z = 860$ pc above
the plane of the edge-on spiral galaxy \ngc 4013. It is powered by a young
stellar cluster that, based on timing constraints, formed {\em in situ} in the
thick disk. The properties of this thick disk \HII\ region and its stars
demonstrate that: a) the thick disk contains at least some dense, even molecular
gas from which the stars formed; b) star formation can proceed in the relatively
low-pressure environments of the thick disk; and c) the origins of the gas in
the thick disk are highly varied, including material expelled from the thin disk
as well as metal poor gas originating in the corona or beyond \citepalias[as
discussed in ][]{howk2018a}.

The formation of hot, massive stars requires a substantial mass of dense, cold
(likely molecular) clouds. While it is often assumed that such cold gas is
confined to the thin disk, there is strong evidence for its presence in the
thick disks of spiral galaxies. \citet{garcia-burillo1999} detected extraplanar
CO filaments to $z \la 300$ pc above the disk of \ngc 4013, demonstrating the
presence of \htwo -bearing gas several scale heights above the plane (with a
total mass of order $\sim10^7$ \msun). The extraplanar CO  may not be
wide-spread, as it was only imaged over the central regions of the galaxy.  Some
of the CO may in fact be associated with a large \halpha\ structure tracing an
apparent outflow from a nuclear starburst \citep{rand1996}.

While some of the CO emission could be associated with rather specific
structures associated with nuclear outflows, we also see dusty thick disk clouds
visible in absorption against the background stellar light over a significant
fraction of the disk. Such clouds are common to spiral galaxies \citep{howk1999,
howk2000}. These clouds are only visible due the contrast they produce,
requiring them to be denser than their surroundings.  Their high densities
($n_{\rm H} \ga 10$ to $\ga25$ \percc) and high \HI\ columns ($\logNHI \ga
10^{20} - 10^{21}$) are consistent with molecular-bearing clouds in the Milky
Way, and the cloud masses approach those of giant molecular clouds
\citep{howk2000, rueff2013}.

Thus, the fuel for star formation exists in the thick disk of \ngc 4013, but
{\em why} does such dense, cold gas exist in the thick disk?  The vast majority
of the gas in the thick disk is thought to have been ejected from the plane,
e.g., by the effects of stellar feedback \citep{norman1989} or through
hydrodynamic effects \citep[see][for an extensive discussion]{howk1997}.
Indeed, the dust seen in broadband optical images (such as the inset of Figure
\ref{fig:image}) almost certainly originated in the thin disk, as the dust
content of infalling or condensing matter is typically low. It may be the case
that the material is lifted from the disk cold, either as the walls of expanding
shells or as material entrained in the outflows. While it seems likely the
transfer of momentum from the hot fluid to the cold clouds would be inefficient,
molecular gas is seen in outflows from starbursts \citep[e.g.,][]{walter2017}.

Even if the gas lifted from the disk is not originally cold/dense enough to
support star formation, the interactions of adjacent outflows in the thick disk
can serve to produce cold material.  Converging flows of neutral material can
compress gas and induce cooling and star formation \citep[e.g.,][]{heitsch2008,
wu2017}; the convergence of the walls of adjacent supershells in the thick disk
may be a trigger for cloud formation \citep{ntormousi2011}. Indeed, the
propensity for outflows to cause the compression of material leading to cold
cloud formation was noted some time ago, whether the outflows were driven by
feedback \citep{de-avillez2000} or by hydraulic jumps at spiral arm shocks
\citep{martos1999}. Gravitational effects, such as tidal perturbations by
passing globular clusters \citep{de-la-fuente-marcos2008} or extended density
waves \citep{struck2009}, could also plausibly lead to triggering cold cloud
formation and star formation in high-altitude gas.

While there is extraplanar star formation in \ngc 4013, it is not so substantial
that it will strongly affect the evolution of the thick disk. If we assume all
the \HII\ region candidates trace star formation (6 EHRs in total with $0.86 \la
z \la 3.4$ kpc), the total \Ha\ luminosity of these candidate \HII\ regions is
$L_{\rm H\alpha , total} \sim 7\times10^{37}$ ergs~s$^{-1}$ (using the MODS
spectra of EHR1 to calibrate our \Ha\ images).  This implies a total thick disk
star formation rate ${\rm SFR} \sim 5\times10^{-4}$ \msunyr\ in \ngc 4013
\citep[using the \Ha -SFR calibration of ][]{kennicutt2012}.

This star formation is also not likely to strongly impact the interstellar thick
disk. One could imagine the presence of young, hot stars in the thick disk could
contribute to the ionization of the DIG, which otherwise requires the escape of
significant ionizing flux from the disk \citep{haffner2009}. If photons produced
by {\em in situ} thick disk stars experience significantly lower opacity, they
could contribute disproportionately to powering the DIG. However, the ionizing
photon budgets typically required for the DIG are typically $\ga 10\%$ of that
produced in the thin disk \citep{haffner2009}, several orders of magnitude
higher than that available from the thick disk star formation in \ngc 4013.

The star forming environments of the thick disk may be similar to those
encountered in the outskirts of galaxies \citep[e.g.,][]{ferguson1998} or in
tidal matter, including the tidal dwarf galaxies \citep[e.g.,][]{werk2008,
de-mello2012}. In these cases, the average gas densities and pressures are
expected to be lower than in the mid-plane of a massive spiral. Additionally, as
is the case with tidally-stripped material, there is not a consistent gravity
vector drawing the clouds together. In order to reach the high densities needed
for star formation, some combination of spontaneous thermal instabilities, tidal
forces (for stripped gas), or converging flows may be required. As more
information becomes available on extraplanar \HII\ regions, a comparison of the
SFRs and \HI +CO columns should allow us to assess global star formation scaling
laws at the lowest SFRs.

There are now a few extraplanar \HII\ regions for which spectroscopic
confirmation has been presented in the literature
\citep[][\citetalias{howk2018a}]{tullmann2003, stein2017}. We summarize their
properties in Table \ref{tab:EHRs}. The \HII\ regions in \citet{stein2017} are
associated with tidally- and ram pressure-stripped gas about spiral galaxies,
reminiscent of those studied in tidal debris by \cite{werk2008}. Two of their
three \HII\ regions show disk-like metallicities, a third is marginally lower.
In \ngc 55 \citep{tullmann2003}, the extraplanar \HII\ regions found at $z \sim
1.1$ and 2.2 kpc have abundances higher than the local disk \citepalias[in the
reassessment of][]{howk2018a}.  This dwarf galaxy shows an incredible wealth of
\halpha\ filaments and loops \citep{ferguson1996, tullmann2003}, and the \HII\
regions may probe gas associated with large-scale outflows from the disk.  The
extraplanar \HII\ regions in \ngc 4013 and \ngc 55 at the very least imply
metallicities of the thick disk are inhomogeneous on relatively small scales.
Future observations will help us assess wehther these changes represent
organized gradients as opposed to stochastic variations within the complex
distribution of gas and dust in the thick disk of this and other galaxies.

The extraplanar \HII\ regions so far reported in the luminous spiral galaxies
all require several O-type stars to provide for their ionization, although there
is almost certainly a selection effect at work given the small number of \HII\
regions so far described in the literature. \ngc 4013 EHR1 has a luminosity as
large as those \HII\ regions forming in tidal debris about \ngc 3628 and \ngc
4522, even though the mechanisms for forming the clusters powering these \HII\
regions are likely quite different. The luminosity of \ngc 55 EHR2\footnote{Only
the luminosity for \ngc 55 EHR2 was reported in \cite{ferguson1996}, which is
the value used by \cite{stein2017}.} is nearly an order of magnitude lower than
those for the brightest in the more luminous spiral galaxies
\citep{ferguson1996}.  This \HII\ region is consistent with being powered by a
single O9.5 or B0 star \citep{stein2017}.

\begin{deluxetable*}{lccccc}
  \tabletypesize{\footnotesize}
  \tablewidth{0pc}
  \tablecolumns{5}
  \tablecaption{Extraplanar \HII\ Region Properties
    \label{tab:EHRs}}
  \tablehead{\colhead{Galaxy} & \colhead{Nebula ID} &
    \colhead{\z\ (pc)} &
    \colhead{\DeltaO} &
    \colhead{$L_{\halpha}$ (ergs s$^{-1}$)} &
    \colhead{Reference}}
  \startdata
  NGC~4013  & 1 & 0.9  & $-0.32\pm 0.09$ & $4 \times 10^{37}$ & This work \\
  NGC~55    & 1 & 1.1  & $+0.24\pm 0.12$ & \nodata & \cite{tullmann2003} \\
  NGC~55    & 2 & 2.2  & $+0.29\pm 0.12$ & $6\times 10^{36}$ & \cite{tullmann2003} \\
  NGC~3628  & 2 & 2.8  & $-0.17\pm 0.10$ & $2\times 10^{37}$ & \cite{stein2017} \\
  NGC~3628  & 3 & 3.0  & $-0.05\pm 0.11$ & $1\times 10^{37}$ & \cite{stein2017} \\
  NGC~4522  & 1 & 1.4  & $-0.03\pm 0.07$ & $6\times 10^{37}$ & \cite{stein2017} \\
  \enddata
  \tablecomments{The nebular IDs are those reported in the original references.
  All abundance offsets \DeltaO\ were derived anew in \citetalias{howk2018a}
  using the intensities reported in the original references. Where more than one
  disk reference region is available (for \ngc 4013 and \ngc 4522), the values
  for \DeltaO\ represent the joint distribution. The \halpha\ luminosity for
  \ngc 55 EHR2 quoted in \cite{stein2017} comes from the images of
  \cite{ferguson1996}. We have corrected the values used in those works to
  reflect the new distance $D = 2.3$ Mpc from \cite{kudritzki2016}.}
\end{deluxetable*}

\section{Summary}
\label{sec:summary}

We have presented observations of a region of the thick disk in \ngc 4013 that
shows strong, point-like \halpha\ emission and a blue continuum source. Based on
ground-based optical spectroscopy and \hst\ imaging, we demonstrate:

\begin{enumerate}

    \item The source EHR1 represents an \HII\ region projected $z = 860$ pc
    above the plane of this galaxy. The ionizing flux required by the \halpha\
    emission is $\approx4 - 7$ times that of the Orion Nebula, requiring the
    equivalent of $\sim6$ O7 V stars.

    \item The optical emission from the underlying continuum source is slightly
    extended, with a FWHM$\, \approx 12$ pc. Thus the object appears to be a
    cluster of stars in the thick disk.

    \item The absolute magnitude and colors of the emission require
    contributions from young stars, the equivalent of $\sim30$ O7 V stars. The
    timescales for the evolution of O stars are inconsistent with the cluster's
    formation in the disk. Thus the star were formed {\em in situ} in the thick
    disk.

    \item EHR1 is by far the \halpha -brightest of the extraplanar \HII\ region
    candidates in \ngc 4013. The total star formation rate of the thick disk in
    this galaxy is quite low, $\sim 5 \times 10^{-4}$ \msunyr. While such a low
    rate of star formation in spiral galaxies will not change the appearance of
    their thick disk populations, its presence allows us to probe the
    metallicity of thick disk gas and to study star formation in unusual
    environments.

\end{enumerate}

\acknowledgements

JCH recognizes the hospitality of the Instituto de Astrof\'{i}sica, Pontificia
Universidad Cat\'{o}lica de Chile during the writing of this work.  We
acknowledge support from NASA through grant NNX10AE87G. Based on data acquired
using the Large Binocular Telescope (LBT).  The LBT is an international
collaboration among institutions in the US, Italy, and Germany. LBT Corporation
partners are the University of Arizona, on behalf of the Arizona university
system; Instituto Nazionale do Astrofisica, Italy; LBT Beteiligungsgesellschaft,
Germany, representing the Max Planck Society, the Astrophysical Institute of
Postdam, and Heidelberg University; Ohio State University, and the Research
Corporation, on behalf of the University of Notre Dame, the University of
Minnesota, and the University of Virginia.  Includes observations obtained with
the NASA/ESA Hubble Space Telescope operated at the Space Telescope Science
Institute, which is operated by the Association of Universities for Research in
Astronomy, Inc., under NASA contract NAS5-26555.

\software{Astropy \citep{price-whelan2018},
  Matplotlib \citep{hunter2007}}

\facilities{LBT(MODS), HST(WFPC2), WIYN}


\bibliography{edge_on_galaxies}


\end{document}